\documentclass[aps,prb,preprint,superscriptaddress,showkeys]{revtex4-2}
\usepackage{graphicx} 
\usepackage{amsmath,amssymb}
\usepackage{hyperref}
\usepackage{bm}

\begin{document}

\title{Quantum-Hall Spectroscopy of Elliptically Deformed Graphene Nanobubble Qubits}

\author{Myung-Chul Jung}
\affiliation{Department of Physics Education, Chosun University, Gwangju 61452, Republic of Korea}
\affiliation{Institute of Well-Aging Medicare \& Chosun University G-LAMP Project Group, Chosun University, Gwangju 61452, Republic of Korea}

\author{Nojoon Myoung}
\email{nmyoung@chosun.ac.kr}
\affiliation{Department of Physics Education, Chosun University, Gwangju 61452, Republic of Korea}
\affiliation{Institute of Well-Aging Medicare \& Chosun University G-LAMP Project Group, Chosun University, Gwangju 61452, Republic of Korea}

\begin{abstract}
    With recent advances in strain-engineering technology of graphene and 2D materials, graphene quantum dots (QDs) defined by the strain-induced pseudo-magnetic fields (PMFs) have been of interest, with the feasibility of tunable graphene qubits\cite{park2023strain}. Here, we theoretically investigate how the electronic states of the nanobubble QDs are influenced by the geometrical anisotropy of the elliptical-shape nanobubbles. We examine the energy levels of the single QD (SQD) and double QD (DQD) spectra by varying the elliptical deformation in the $x$ and $y$ axes, respectively. We found that the SQD and DQD show distinguished behavior with respect to the direction of the elliptical deformation. While the SQD levels are substantially affected by the $y$-directional deformation, the DQD levels are largely shifted by the $x$-directional deformation.
\end{abstract}


\maketitle

\section{Introduction}

Graphene has emerged as a promising platform for quantum information technologies because its charge, spin, and valley degrees of freedom exhibit long relaxation and coherence times\cite{oh2010electronic,volk2013probing,banszerus2022spin,gachter2022single,garreis2024long}. However, the renowned Klein tunneling---arising from chirality-protected suppression of backscattering\cite{young2009quantum,allain2011klein}---makes it difficult to control Dirac fermions with the electrostatic potentials routinely employed in metal-oxide-semiconductor devices. Although quantum-dot (QD) architectures have been pursued as a route to graphene qubits\cite{recher2010quantum,shafraniuk2020graphene,garreis2024long,banszerus2022spin,wu2012graphene,banszerus2020electron,park2023strain}, the absence of an intrinsic band gap remains a central obstacle to electrical confinement. This limitation motivates alternative strategies---strain-induced pseudo-magnetic fields(PMFs)---that can localize Dirac fermions without sacrificing their exceptional coherence. 

Strain engineering has therefore become a possible route to tailor the electronic landscape of graphene. Because monolayer graphene can sustain elastic deformations approaching 20 $\%$\cite{cao2020elastic,si2016strain,mendoza2021strain}, controlled strain provides a means to tune band structure and manipulate valley degree of freedom\cite{pereira2009strain,guinea2012strain,sahalianov2019straintronics,oliva2013understanding,cocco2010gap,myoung2014gate,myoung2020manipulation,myoung2022strain,jun2025nanowrinkle}. In typical dry-transfer fabrication, graphene nanobubbles arise naturally when nanoscale contaminants are trapped between graphene and its supporting substrate\cite{ghorbanfekr2017dependence}. The seminal scanning-tunneling study by Levy \textit{et al.} revealed Landau-level quantization within such nanobubbles, confirming that strain-generated PMFs can confine Dirac fermions\cite{levy2010strain}. Subsequent theoretical and experimental advances have extended this concept, demonstrating strain-induced flat bands, valley filtering, and programmable PMFs in graphene devices\cite{meng2013strain,liu2022realizing,liu2022analytic,nigge2019room,le2021modulation,wagner2022landau,li2020valley}. These advances now enable QD architectures defined by strain engineering.

PMFs generated by elastic strain in graphene provide a means to tune quantum-transport properties of graphene, and previous work has shown that strain-defined nanobubbles act as QDs in which Dirac fermions are spatially localized with the valley polarization by PMF\cite{myoung2020manipulation}. Strain-engineered graphene QDs offer two key advantages. (i) A local and inhomogeneous PMF confines Dirac fermions without opening a band gap, thereby preserving high coherence. (ii) The quantum states of such strain-induced QDs can be tuned adiabatically by modifying the nanobubble geometry. Advances in scanning-probe and transfer techniques now allow for the reshaping of graphene strain fields and in-situ control of nanobubble aspect ratios\cite{georgi2017tuning,jiang2017visualizing,hsu2020nanoscale}. 

A recent study has proposed strain-induced graphene nanobubble QDs as promising solid-state qubits\cite{park2023strain}. In these systems, Dirac fermion confinement is mediated by the characteristic PMF---a mechanism explored in centro-symmetric (circular) nanobubbles\cite{schneider2015local,moldovan2013electronic,torres2018tuning,myoung2020manipulation,park2023strain2}. Theoretically, graphene nanobubbles adopt circular profiles at equilibrium; however, the nanobubbles observed in practical devices are frequently anisotropic, reflecting substrate roughness or transfer-induced wrinkles\cite{georgi2017tuning,hsu2020nanoscale,ghorbanfekr2017dependence,levy2010strain,palinkas2016determination}. Because the nanobubble qubit's operating parameters are governed by its energy spectra of QDs and inter-dot tunnel coupling---both dependent on bubble geometry---it is crucial to determine how the non-centro-symmetric nanobubble modifies qubit energetics relative to ideal circular counterparts.

In this work, we model geometrically deformed nanobubbles with a Gaussian height profile whose cross-section is elliptically given. By varying the ellipticity along the $x$ and $y$ axes, we compute how the energy levels of the nanobubble QDs shift. For double-quantum-dot (DQD) configurations, we determine the two-level energy splitting that defines the qubit energy gap, and show that this gap depends on geometric anisotropy. The qubit energy increases when the nanobubble is elongated along the $x$-axis, but decreases for elongation along the $y$-axis, an opposite directional dependence rooted in the anisotropic PMF profile. This pronounced, geometry-driven tunability demonstrates that strain engineering of nanobubble shape can serve as a programmable control knob for graphene-based qubits, advancing the prospect of graphene straintronics for quantum information technologies.

\section{Model and Methods}

In this section, we present the theoretical framework used to model geometrically deformed graphene nanobubbles and to compute quantum transport to probe their qubit energy levels. The elliptical nanobubbles are described within continuum elasticity, which is valid for naturally formed nanobubbles whose height profiles span hundreds of lattice constants. From this displacement field, we derive an analytic expression of the strain-induced PMF, which provides insight into the electronic spectrum of the nanobubble QDs. Meanwhile, the presence of the PMF is also incorporated into a tight-binding Hamiltonian and solved numerically to obtain transport characteristics. Ballistic conductance is calculated with the S-matrix and Landauer-B\"{u}ttiker formalisms, implemented using the \textsc{kwant} \cite{groth2014kwant} and MUMPS\cite{amestoy2001fully} packages.

\begin{figure*}[hbpt!]
    \centering
    \includegraphics[width=0.9\linewidth]{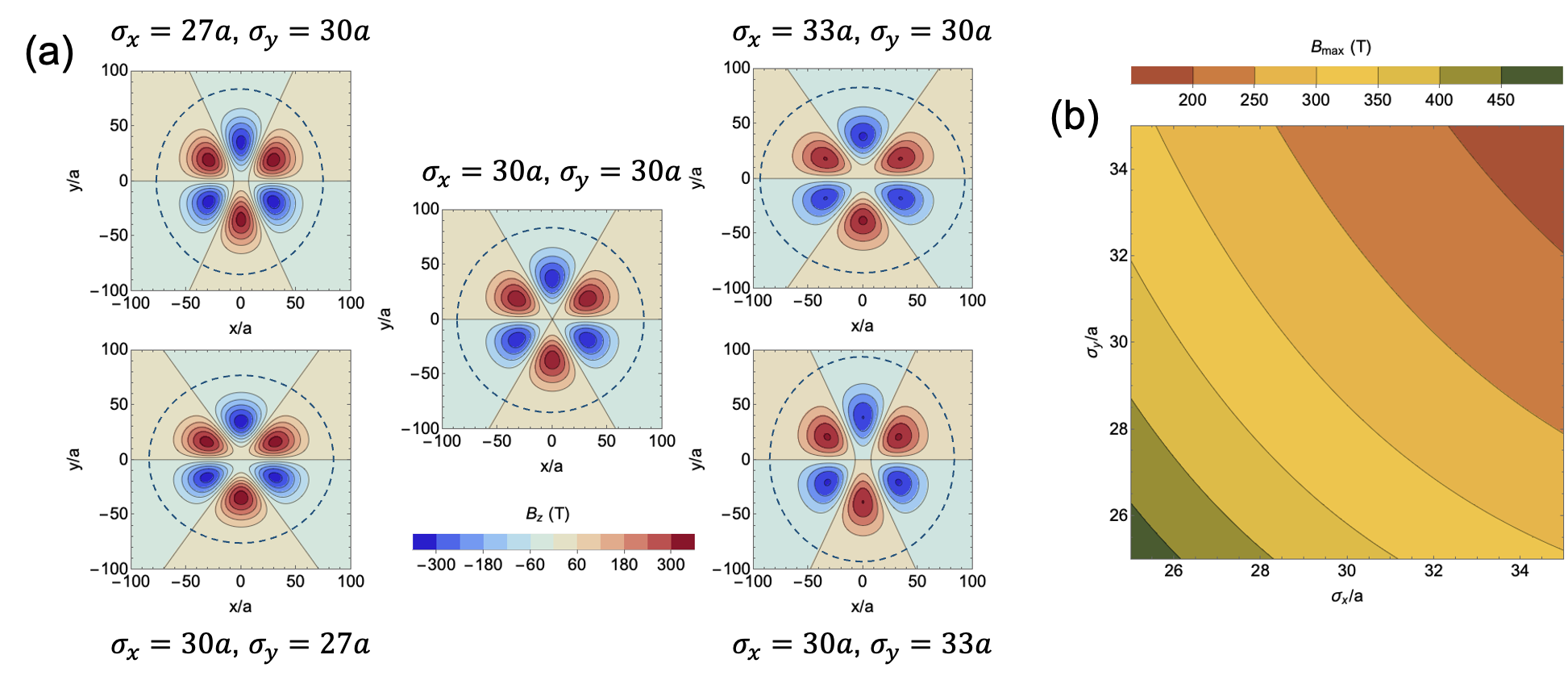}
    \caption{(a) Spatial distribution of the pseudo-magnetic field (PMF) for representative elliptic nanobubbles for various $\sigma_{x}$ and $\sigma_{y}$ values. All plots are displayed with the same PMF strength range, $-360~\mathrm{T}<B_{z}<+360~\mathrm{T}$. (b) The maximum PMF strength versus $\sigma_{x}$ and $\sigma_{y}$, for $h_{0}=17a$.}
    \label{fig:model}
\end{figure*}

We model the vertical deformation of graphene nanobubbles by an elliptical Gaussian height profile
\begin{align}
    h\left(x,y\right)=h_{0}~\mathrm{exp}{\left\{-\frac{1}{2}\left[\left(\frac{x}{\sigma_{x}}\right)^{2}+\left(\frac{y}{\sigma_{y}}\right)^{2}\right]\right\}},
\end{align}
where $h_{0}$ is the maximum height of the nanobubble and $\sigma_{x,y}$ represents the lateral widths in $x,y$ directions, respectively. Neglecting in-plane displacements, the strain tensor $u_{ij}=\left(\partial_{i}h\right)\left(\partial_{j}h\right)/2$, which generates a valley-dependent gauge field
\begin{align}
    A_{x}=\frac{\nu\sqrt{3}\hbar\beta}{ea}\left(u_{xx}-u_{yy}\right),\qquad A_{y}=-2\frac{\nu\sqrt{3}\hbar\beta}{ea}u_{xy},
\end{align}
where $\nu=\pm$ labels the $K/K'$ valleys, $a=0.246~\mathrm{nm}$ is the lattice constant of graphene and $\beta=3.37$. The PMF, $\vec{B}=\vec{\nabla}\times\vec{A}$, becomes
\begin{align}
    \vec{B}= &\frac{\nu\sqrt{3}\hbar \beta}{ea}\frac{h_{0}^{2}}{\sigma_{x}^{4}\sigma_{y}^{6}}\left[\left(-y^{2}+\sigma_{y}^{2}\right)\sigma_{x}^{4}+\left(3x^{2}-\sigma_{x}^{2}\right)\sigma_{y}^{4}\right]\nonumber\\
    &\times\mathrm{exp}\left[-\left(\frac{x}{\sigma_{x}}\right)^{2}-\left(\frac{y}{\sigma_{y}}\right)^{2}\right]\hat{z}.
\end{align}
Here, let us mention that $\vec{B}$ no longer obeys the $\mathcal{C}_{3v}$ symmetry, contrary to that of circular nanobubbles. It is also worth noting that the maximum strength of the PMF varies with the ellipse parameters, $\sigma_{x}$ and $\sigma_{y}$, as shown in Fig. \ref{fig:model}.

The electronic system of the quantum Hall graphene hosting both the nanobubble and a gate-defined p-n interface is described by the tight-binding Hamiltonian
\begin{align}
    H=&t_{0}\sum_{\left<i,j\right>}e^{-\beta\left(\frac{d_{ij}}{a_{0}}-1\right)}e^{-i\frac{\varphi}{2}\left(x_{i}-x_{j}\right)\left(y_{i}+y_{j}\right)}+U\left(\vec{r}_{i}\right)
    \label{eq:Ham}
\end{align}
where $t_{0}=2.7~\mathrm{eV}$ is the unstrained hopping amplitude, $a_{0}=a/\sqrt{3}=0.142~\mathrm{nm}$ is the carbon-carbon distance of pristine graphene, $d_{ij}$ is the strain-modified bond length, and $\vec{r}_{i}=\left(x_{i},y_{i}\right)$ is the position of an $i$-th atomic site of the system. $\varphi=\Phi/\Phi_{0}$ is the magnetic flux through a unit cell per a flux quantum $\Phi_{0}=h/e$, and we adopt the Landau gauge $\vec{A}_{ext}=\left(-B_{0}y,0,0\right)$, which produces a uniform external field $\vec{B}=\left(0,0,B_{0}\right)$. The on-site energy $U\left(\vec{r}_{i}\right)$ models the electrostatic potential of the p-n junction.

Quantum-Hall channels formed along a graphene p-n junction provide a coherent beam splitter that has enabled electronic Mach-Zehnder interferometers\cite{jo2022scaling,jo2021quantum,wei2017mach,mirzakhani2023electronic,myoung2024detecting,myoung2017conductance}. These same interface channels have also been exploited as spectroscopic probes of localized states in strain-induced graphene nanobubbles\cite{myoung2020manipulation,park2023strain,myoung2024detecting}. Here, we use the quantum-Hall interface channels to quantify how the discrete energy levels of nanobubble quantum dots evolve under controlled geometric deformation of the nanobubble, thereby linking transport signatures to the anisotropic PMF analyzed in the previous subsection.

At a certain Fermi energy of the system, mode-resolved transmission probabilities $T_{nm}\left(E\right)$ between mode $m$ in lead $\beta$ and mode $n$ in lead $\alpha$ are calculated with the S-matrix formalism, and the ballistic conductance is formulated using Landuer-B\"{u}ttiker approach
\begin{align}
    G\left(E\right)=\frac{2e^{2}}{h}\sum_{n\in\alpha,m\in\beta}T_{nm}\left(E\right).
\end{align}
Lastly, note that the ballistic transport regime is secured when the magnetic length $l_{B}=\sqrt{\hbar/2eB_{0}}$ of the system is sufficiently shorter than the system size. Throughout this study, the system size is in units of hundreds of nanometers, and we have $l_{B}=12a=2.96~\mathrm{nm}$ for $B_{0}=37.7~\mathrm{T}$, with the tight-binding flux parameter $\varphi=0.003$ in Eq. (\ref{eq:Ham}). Thus, we guarantee that the quantum transport phenomena presented in this paper occur along the quantum-Hall channels.

\section{Results and Discussion}

\subsection{Deformation effects on elliptical nanobubble QDs}

\begin{figure*}[htpb!]
    \centering
    \includegraphics[width=\linewidth]{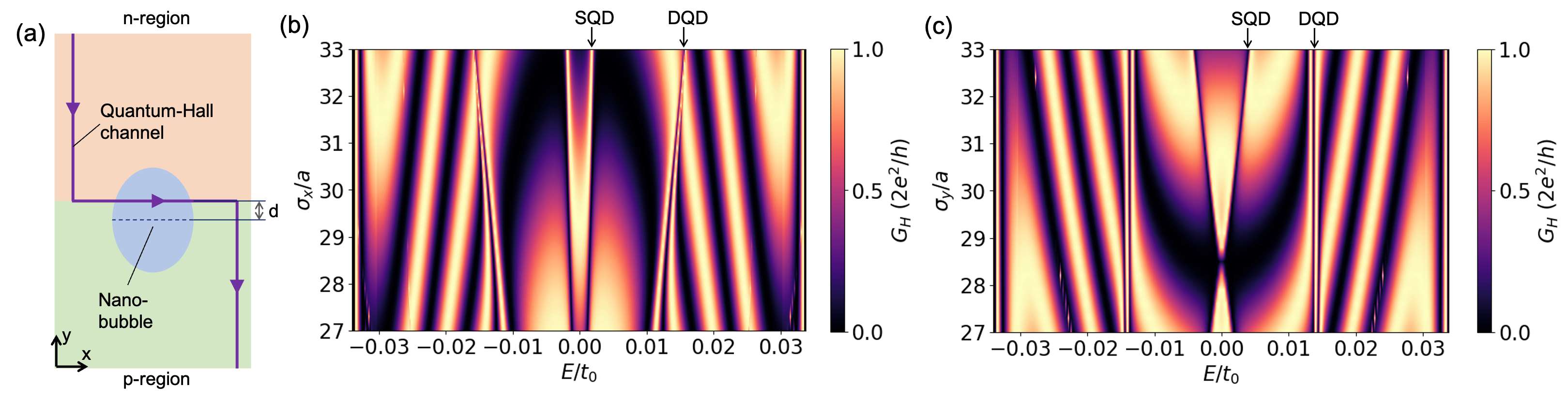}
    \caption{(a) Schematics of quantum-Hall spectroscopy used to probe the localized states of the nanobubble QD. The two sides of the p-n junction are tuned to filling factors $\nu=\pm1$ by the anti-symmetric potential step defined in Eq. (\ref{eq:pnjnc}). The distance between the quantum-Hall channel and the nanobubble center is denoted by $d$. (b) Color map of the quantum-Hall conductance $G_{H}$ as a function of Dirac fermion energy $E$ and deformation parameter $\sigma_{x}$ for fixed $\sigma_{y}=30a$. (c) Analogous map $G_{H}$ versus $E$ and $\sigma_{y}$ for fixed $\sigma_{x}=30a$. Fano resonances arising from SQD and DQD bound states are indicated by solid arrows. A spectra are calculated with $h_{0}=17a$.}
    \label{fig:QDlevels-energy}
\end{figure*}

Localized states in the nanobubble QDs are identified through Fano resonances that appear in the quantum-Hall conductance $G_{H}$ measured via the quantum-Hall interface channel of a graphene p-n junction [Fig. \ref{fig:QDlevels-energy}(a)]. Fano resonance arises from quantum interference between the discrete bound states of the nanobubble QD and the extended quantum-Hall modes, producing the characteristic asymmetric line shapes. The electrostatic profile of the p-n junction is modeled as
\begin{align}
    U\left(y\right)=\frac{U_{0}}{2}\tanh{\left(\frac{y}{\psi}\right)}, \label{eq:pnjnc}
\end{align}
where $E_{0}=\sqrt{2\hbar v_{F}eB_{0}}$ is the separation between the zeroth and first Landau levels in graphene under a homogeneous external magnetic field $B_{0}$, and $\psi=30a$ is a parameter setting the smoothness of the p-n junction. We investigate how Fano resonances evolve as the nanobubble deformation parameters $\left(\sigma_{x},\sigma_{y}\right)$ are varied, thereby quantifying the impact of ellipticity on the QD energy spectrum.

Figures \ref{fig:QDlevels-energy}(b) and (c) display the conductance maps $G_{H}$, obtained by varying $\sigma_{x}$ for fixed $\sigma_{y}=30a$ and varying $\sigma_{y}$ for fixed $\sigma_{x}=30a$, respectively. In the energy window $-E_{0}<E<E_{0}$, two well-distinct Fano resonances appear, corresponding to single quantum-dot (SQD) and the DQD bound previously defined in circular nanobubbles\cite{park2023strain}. From the results, we find the opposite response of the SQD and DQD levels to elliptical deformation:
\begin{itemize}
    \item $x$-direction ($\sigma_{x}$) deformation : The SQD resonance shifts only slightly, whereas the DQD resonance exhibits a pronounced change---both its linewidth and the two-level splitting decrease markedly as $\sigma_{x}$ decreases.
    \item $y$-direction ($\sigma_{y}$) deformation : Changed $\sigma_{y}$ produces a substantial energy shift in the SQD resonance, while the DQD resonance remains almost insensitive in both energy and linewidth.
\end{itemize}
This complementary behavior highlights the anisotropic pseudo-magnetic confinement and inter-dot coupling.

The opposite response of the SQD and DQD resonances can be understood from the distinct spatial localization of their bound states. A previous study showed that strained-induced bound states in a nanobubbles are preferentially at the positions where the PMF reaches its extrema\cite{myoung2020manipulation}. Denoting by $r_{0}$ the radial distance of these hot spots from the nanobubble center, a circular nanobubble ($\sigma_{x}=\sigma_{y}$) hosts SQD bound states at $\left(x_{S},y_{S}\right)=r_{0}\left(0,1\right)$, and DQD bound states at $\left(x_{D},y_{D}\right)=r_{0}\left(\pm\sqrt{3}/2,1/2\right)$. When the nanobubble is stretched along the $y$-axis (increasing $\sigma_{y}$), the field extrema and hence the SQD bound states move predominantly along $y$, producing a sizable shift of the SQD levels while leaving the DQD level relatively intact. In contrast, increasing $\sigma_{x}$ displaces the DQD hot spots along $x$, strongly modulating their tunnel coupling, whereas the SQD levels (on the minor axis of the ellipse) are weakly affected. This geometric selectivity explains why the SQD resonance is more sensitive to $\sigma_{y}$ deformation, whereas the DQD resonances exhibit a noticeable response to $\sigma_{x}$ deformation. 

Because the quantum-Hall channels along a p-n junction can be displaced continuously by tuning the Fermi energy\cite{handschin2017giant,myoung2017conductance}, the conductance maps in Fig. \ref{fig:QDlevels-energy} encode not only the energetic shift of the nanobubble QD levels, but also the concomitant spatial displacement of the QDs' localized states induced by elliptical deformation. Precise knowledge of this real-space location is essential for gate-defined qubit modulation, which requires sophisticated device fabrication. We therefore analyze Fano resonances in $G_{H}$ spectra by varying the relative position between the nanobubble and the p-n junction interface, as shown in Fig. \ref{fig:QDlevels-energy}(a).

\begin{figure}[htpb!]
    \centering
    \includegraphics[width=0.5\linewidth]{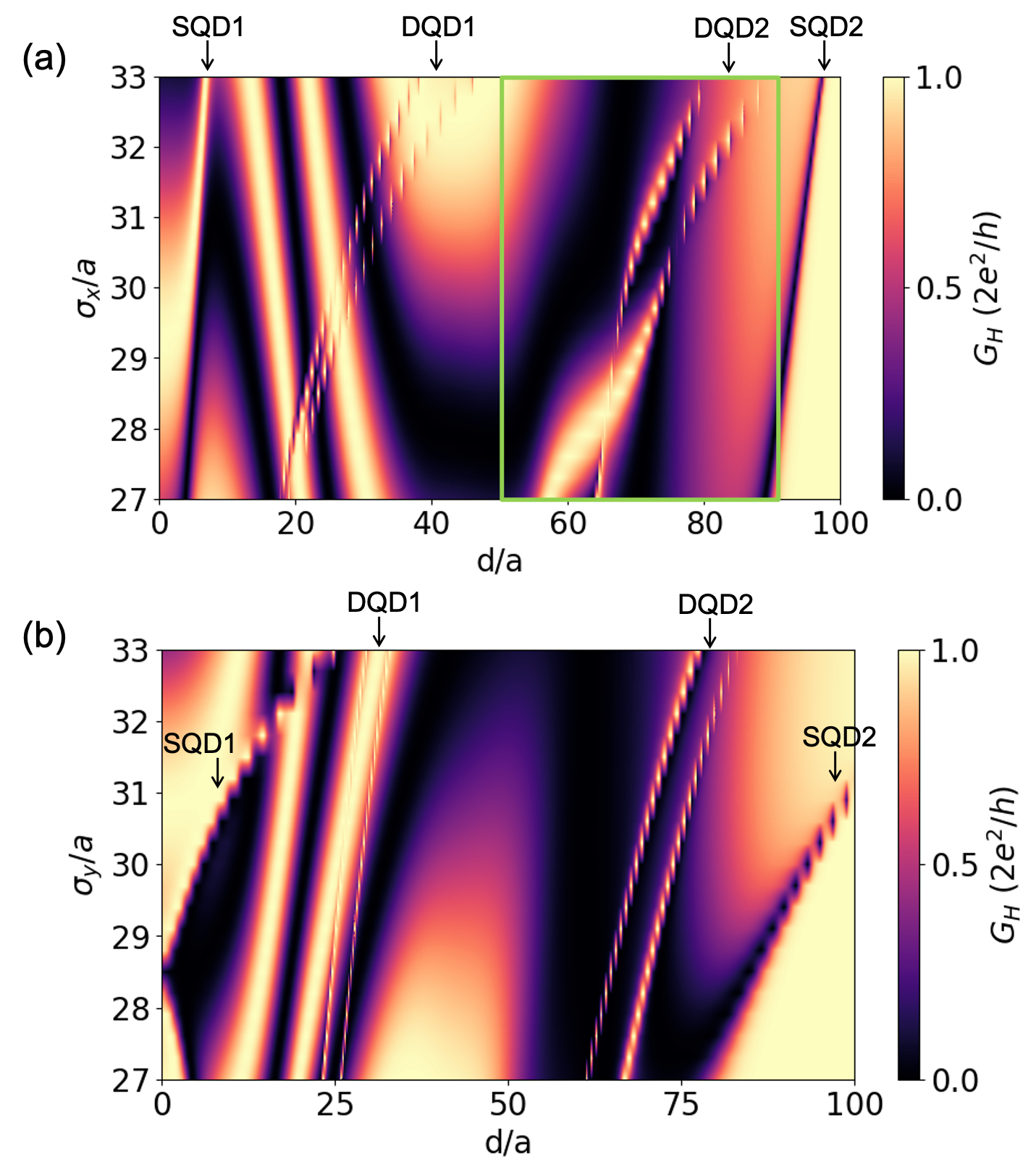}
    \caption{(a) Quantum-Hall conductance maps showing how the resonances evolve as the relative distance between the quantum-Hall channels and the nanobubble, $d$, and $\sigma_{x}$; the minor-axis width is fixed at $\sigma_{y}=30a$. (b) Analogous data obtained by varying $\sigma_{y}$ at fixed $\sigma_{x}=30a$. The primary and secondary SQD resonances are labeled SQD1 and SQD2; the corresponding DQD resonances are labeled DQD1, DQD2.}
    \label{fig:QDlevels-spatial}
\end{figure}

Figure \ref{fig:QDlevels-spatial} shows how the bound states of the elliptical nanobubble QD move in real space by varying the deformation parameters, $\sigma_{x}$ and $\sigma_{y}$. When nanobubbles are stretched along the $x$-axis (increasing $\sigma_{x}$), the DQD levels shift far more than the SQD levels, reflecting the strong sensitivity of the DQD pair to changes in the major axis. In contrast, the $y$-axis stretching (increasing $\sigma_{y}$) produces a pronounced shift in the SQD levels, while a shift in the DQD levels is found to be relatively smaller. In both cases, the localized states move outward, away from the nanobubble center, as the corresponding lateral size of the nanobubble is increased, confirming that the geometric deformation directly relates to the spatial registry of the nanobubble qubit states.

\begin{figure*}[htpb!]
    \centering
    \includegraphics[width=\linewidth]{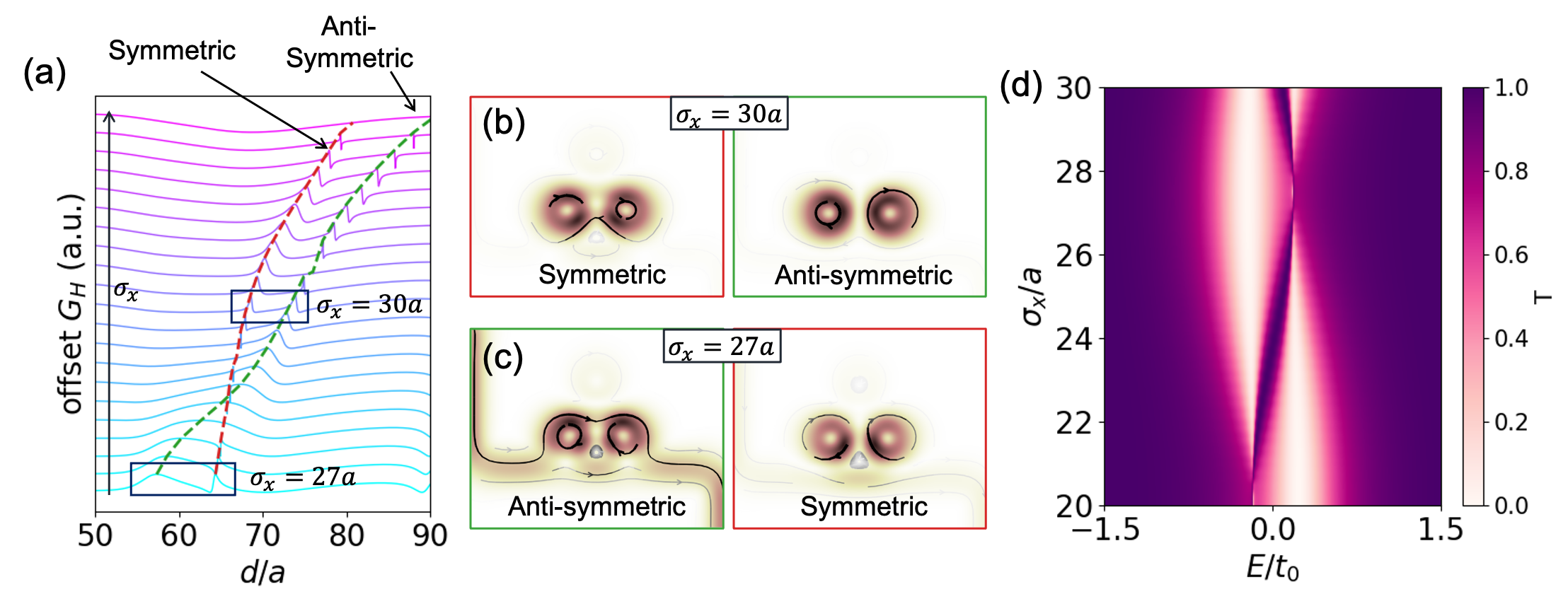}
    \caption{(a) Fano-resonance switch for the DQD2 states in the $Q_{H}$ spectra corresponding to the highlighted area in Fig. \ref{fig:QDlevels-spatial}(a). Green and red dashed lines represent the Fano-resonance energies for the anti-symmetric and symmetric configurations of the DQD2, respectively. (b) and (c) probability current density maps corresponding to the Fano resonances for the circular and elliptical nanobubble DQD2. (d) Calculated results of the reduced transport model through the side-coupled DQD with complex inter-dot hoppings. Here, we set $h_{0}=17a$ and $\sigma_{y}=30a$.}
    \label{fig:Fanoshift}
\end{figure*}

\subsection{Strain-mediated phase shift in DQD states}

Interestingly, the secondary DQD levels, labeled DQD2, show a pronounced level crossing as $\sigma_{x}$ is varied [green box in Fig. \ref{fig:QDlevels-spatial}(a)]. Because graphene has a universal Berry phase of $\pi$\cite{zhang2005experimental}, the ground-state DQD1 levels are anti-symmetric, whereas the excited DQD1 levels are symmetric. Figure. \ref{fig:Fanoshift} shows that this level configuration is reversed for the DQD2 branch when $\sigma_{x}$ is changed, indicating a strain-mediated phase shift in addition to the universal Berry phase (detailed derivation in App. \ref{app:Berryphase}).

We attribute the crossing to a complex inter-dot hopping amplitude mediated by the PMF when a third QD state becomes weakly occupied. A similar Berry-phase switch driven by complex tunneling was reported in a gate-tunable three-level system of semiconductor QDs\cite{wang2008voltage}. To capture the present phenomenon, we construct a reduced transport model consisting of a one-dimensional quantum-Hall channel coupled to DQD inside the nanobubble (details in App. \ref{app:Berryphase}). The calculated conductance map, Fig. \ref{fig:Fanoshift}(d), reproduces the Fano-resonance inversion observed when $\sigma_{x}$ is compressed, provided that the hopping amplitude acquires the phase generated by the PMF flux enclosed by the inter-dot pathways of a three-QD case.

Figures \ref{fig:Fanoshift}(b) and (c) further reveal that the third QD is appreciably excited when $\sigma_{x}<30a$; concomitantly, the Berry phase of the DQD manifold is tuned. We therefore conclude that strain-controlled phase shift is achievable when an effective triple QD configuration is established within the nanobubble.

\subsection{Energy splitting of elliptic-nanobubble qubits}

So far, we have examined how the elliptical deformation alters the energy spectrum and the real-space location of nanobbuble QDs. We now turn to the quantity that defines their usefulness as qubits: the energy splitting $\Delta$ between the symmetric and anti-symmetric states of a DQD formed when two QD states are maximally superposed. As shown in earlier work\cite{park2023strain}, the splitting sets the qubit transition frequency and can be tuned by controlling the strain.

\begin{figure*}
    \centering
    \includegraphics[width=\linewidth]{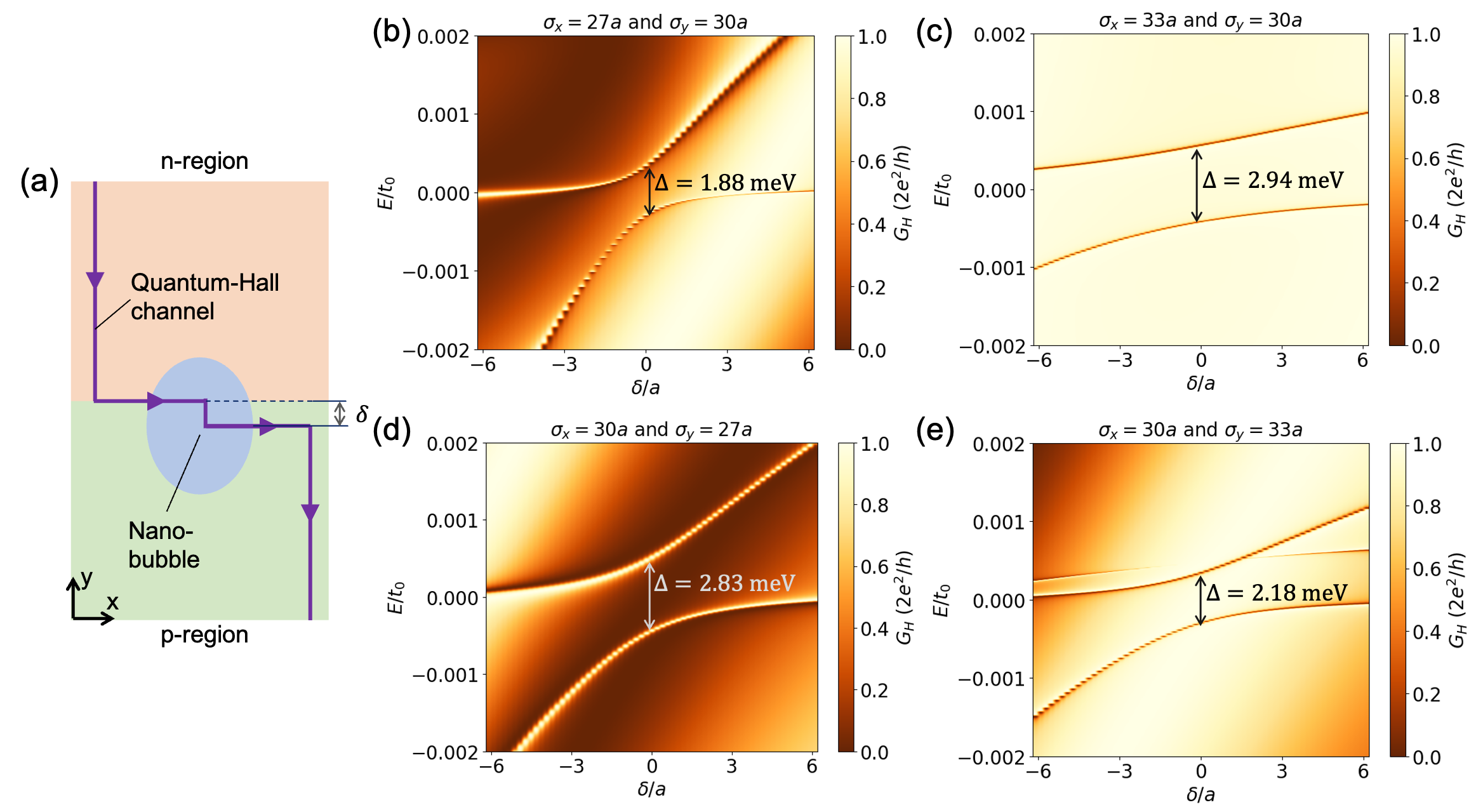}
    \caption{(a) Schematics of the gate-defined modulation of a graphene nanobubble qubit. The position of the quantum-Hall channel is shifted relative to the nanobubble by an electrostatic detuning parameter $\delta$. (b-e) $Q_{H}$ maps for the DQD1 branch, showing the characteristic avoided crossing that defines the qubit energy gap under four representative elliptic deformations: (b) $\sigma_{x}=27a$, $\sigma_{y}=30a$; (c) $\sigma_{x}=33a$, $\sigma_{y}=30a$; (d) $\sigma_{x}=30a$, $\sigma_{y}=27a$; (e) $\sigma_{x}=30a$, $\sigma_{y}=33a$. The energy splitting $\Delta$ at the anti-crossing corresponding to the qubit transition frequency when the two QDs are maximally superposed.}
    \label{fig:qubits}
\end{figure*}

To bring the two localized states into Fano resonances, we apply an electrostatic detuning generated by a pair of local gates---an approach routinely used in low-dimensional quantum devices. The p-n junction potential is modified to
\begin{align}
    U\left(x,y\right)=\frac{U_{0}}{2}\tanh{\left[\frac{y-y_{0}\left(x\right)}{\psi}\right]},
\end{align}
where $y_{0}\left(x\right)=\delta \left(1+\tanh{x}\right)/2$ and the parameter $\delta$ defines the detuning.

For $\delta=0$, the quantum-Hall channel is straight, so both QDs are equally coupled to the extended quantum-Hall states. A finite $\delta$ bends the quantum-Hall channel, leaving the coupling to one QD unchanged while reducing---or enhancing---that to the other, as sketched in Fig. \ref{fig:qubits}(a).

Figures \ref{fig:qubits}(b-e) plot $G_{H}$ for four representative elliptic deformations. In all cases, clear avoided crossings appear as $\delta$ sweeps through zero, confirming the formation of a two-level system. The gap $\Delta$ at the anti-crossing---corresponding to the qubit transition energy---varies systematically with the deformation parameters $\sigma_{x}$ and $\sigma_{y}$, demonstrating that strain engineering provides a geometric knob for tuning the qubit frequency.

Earlier work reported an energy splitting of $\Delta=2.4~\mathrm{meV}$ for a circular nanobubble qubit ($\sigma_{x}=\sigma_{y}$)\cite{park2023strain}. Our results reveal a clear anisotropic trend: $\Delta$ increases with increasing $\sigma_{x}$, whereas $\Delta$ decreases with increasing $\sigma_{y}$. Hence, the qubit energy scales with the aspect ratio $\sigma_{x}/\sigma_{y}$: it is larger when the nanobubble is stretched along the $x$ direction ($\sigma_{x}>\sigma_{y}$) and smaller when stretched along the $y$ direction.

The deformation also affects the linewidth of the Fano resonances, which gauges how strongly the QD couples to the quantum-Hall extended states. For both types of deformation, the Fano resonances become sharper, indicating weaker coupling as the overall nanobubble size grows. Taken together, these results demonstrate that strain engineering offers control over graphene nanobubble qubits: the transition energy $\Delta$ can be tuned via the aspect ratio of the nanobubble, while the qubit-environment coupling can be adjusted through the nanobubble's lateral dimensions.

\section{Conclusions}

We have analyzed how geometric deformation alters the electronic states of strain-induced pseudo-magnetic field (PMF) quantum dots (QDs) in a graphene nanobubble. The nanobubble was modeled as an elliptic Gaussian height profile so that the strain could be varied independently along the $x$ and $y$ axes. Quantum-Hall channel spectroscopy reveals two distinct sets of bound states---single QD (SQD) and double QD (DQD)---that respond in complementary ways to anisotropic strain:
\begin{itemize}
    \item SQD levels shift substantially when the nanobubble is stretched in the $y$ direction but relatively weakly for the $\sigma_{x}$ deformation.
    \item DQD levels are relatively insensitive to the $\sigma_{y}$ deformation but exhibit pronounced energy and linewidth changes when the nanobubble is deformed along the $x$ direction.
\end{itemize}
This directional selectivity follows from the spatial location of the SQD and DQD bound-state maxima relative to the PMF extrema. When a third QD state becomes excited, the PMF also imparts a complex tunneling amplitude that drives an additional phase shift, producing a symmetry-switching level crossing within the DQD manifold. 

In addition, we have shown that the qubit energy gap $\Delta$ (the avoided-crossing splitting) scales with the nanobubble's aspect ratio $\sigma_{x}/\sigma_{y}$: $\Delta$ increases for $\sigma_{x}>\sigma_{y}$ and decreases in the opposite case. Larger nanobubbles simultaneously sharpen the Fano resonances, indicating weaker coupling between the QDs and the quantum-Hall channel.

Together, these results demonstrate that elliptically deformed nanobubbles provide mechanical control over both the transition energy and relaxation of graphene nanobubble qubits, pointing toward strain-programmable quantum devices in a monolayer graphene platform.

\section*{Data Availability}
The data that support the findings of this study are available from the corresponding author upon reasonable request.

\section*{Acknowledgment}
    This work is supported by a research fund from Chosun University (2022).

\section*{Conflict of Interest}
The authors declare no competing financial interests.

\appendix

\section{Reduced Model for Double Quantum Dot Transport} \label{app:Berryphase}

\begin{figure}[htpb!]
    \centering
    \includegraphics[width=0.5\linewidth]{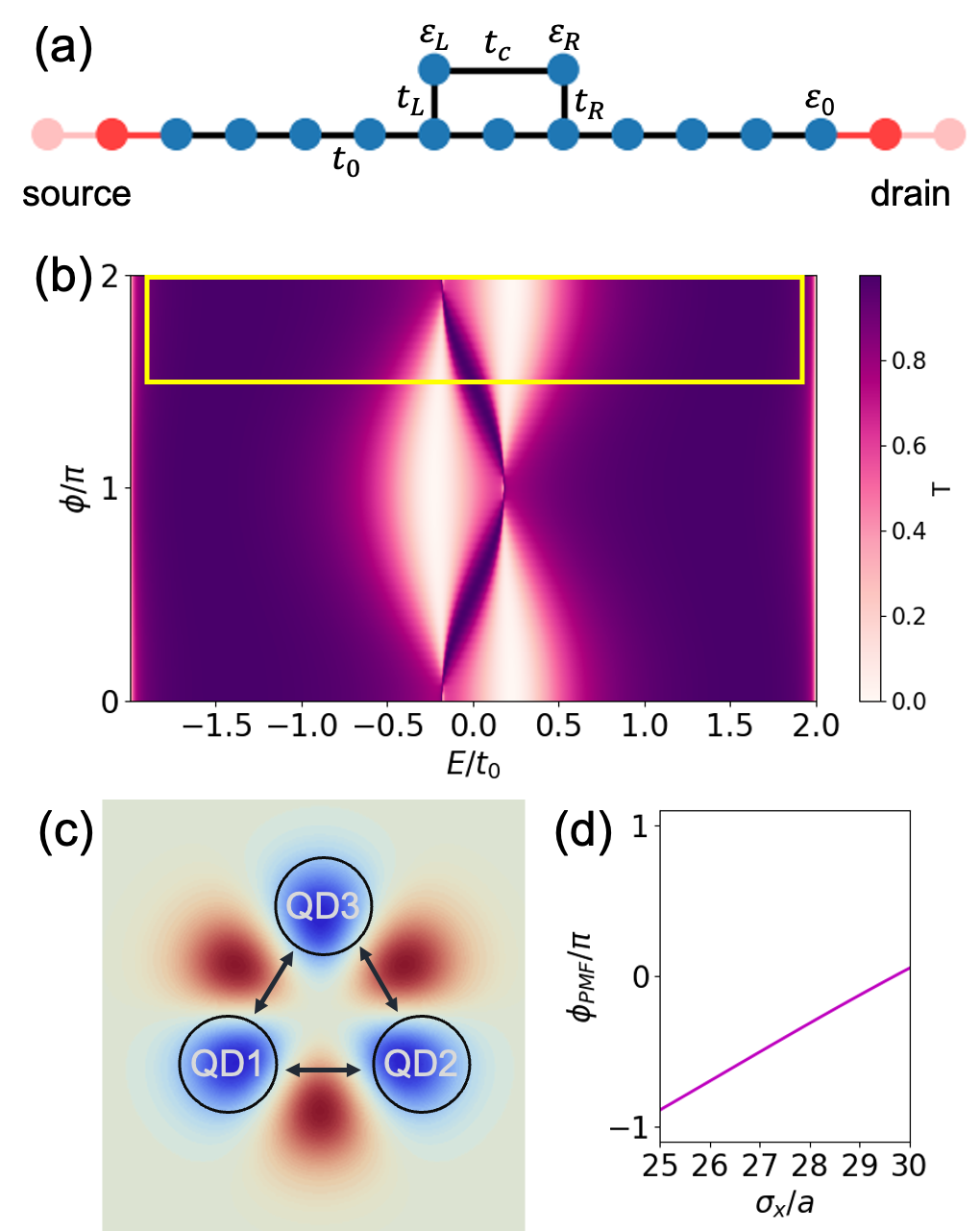}
    \caption{(a) Reduced model for electron transport, mimicking the DQD Fano resonances via the quantum-Hall channel. $\epsilon_{0}$, $\epsilon_{L/R}$ are the on-site energies of the 1D chain and the left/right dots. $t_{0}$, $t_{L/R}$ are hopping amplitudes to describe the 1D chain hopping and the coupling between the chain and the left/right dots. Note that the interaction between two dots is considered as a complex hopping of which amplitude and phase are $t_{c}$ and $\phi$. (b) Transmission spectra versus the electron energy $E$ and $\phi$. (c) Schematic diagram for possible hopping processes (black arrows) of a triple QD, enclosing a triangular area where the PMF penetrates. (d) Calculated $\phi$ due to the PMF-induced flux as a function of $\sigma_{x}$ for a given $\sigma_{y}=30a$.}
    \label{fig:reducedmodel}
\end{figure}

In this work, quantum-Hall spectroscopy of electronic properties of nanobubble QDs is investigated by means of Fano resonances between the quantum-Hall channel and the QDs. To find the physical origin of the Fano-resonance switching effects, we now consider a reduced transport model to mimic the transport phenomena observed in the main text. The reduced model basically consists of a 1D chain and two dots, corresponding to the tight-binding approximation, as shown in Fig. \ref{fig:reducedmodel}(a). The reduced Hamiltonian of the system is developed via the tight-binding approximation:
\begin{align}
    H_{reduced}=H_{chain}+H_{dot}+H_{coupling},
\end{align}
where
\begin{align}
    H_{chain}&=\varepsilon_{0}\sum_{i}a_{i}^{\dagger}a_{i}+t_{0}\sum_{\left<i,j\right>}a_{i}^{\dagger}a_{j}+h.c.,\\
    H_{dot}&=\epsilon_{L}d_{L}^{\dagger}d_{L}+\epsilon_{R}d_{R}^{\dagger}d_{R}+h.c.,\\
    H_{coupling}&=t_{L}d_{L}^{\dagger}c_{4}+t_{R}d_{R}^{\dagger}c_{6}+t_{c}e^{i\phi}d_{L}^{\dagger}d_{R}+h.c.,
\end{align}
where $a_{i}/a_{i}^{\dagger}$ are annihilation/creation operators of the 1D chain, $\varepsilon_{0}$ and $t_{0}$ are the on-site and hopping energies of the 1D chain, $d_{L,R}/d_{L,R}^{\dagger}$ are annihilation/creation operators of the left/right dots, and $\epsilon_{L,R}$ are the dot levels. Each dot is connected to the 1D chain by couplings $t_{L,R}$. In addition, we consider a complex coupling term between two dots with hopping amplitude $t_{c}$ and phase $\phi$. The complex coupling of the DQDs in this work can be introduced by the PMF flux through an area enclosed by pathways between the DQDs and the third QD. In other words, $\phi$ vanishes when the third QD is not excited.

After attaching the source and drain leads to the system, we can compute the transmission probability between two leads, as functions of the electron energy $E$ and the phase $\phi$. The resulting transmission spectra are presented in Fig. \ref{fig:reducedmodel}(b), apparently indicating the level-crossing nature of the anti-symmetric and symmetric DQD states.

To mimic the Berry-phase switching observed in the elliptical nanobubble DQDs under the $\sigma_{x}$ deformation, we implemented the PMF-induced flux to the phase term of the complex inter-QD coupling by considering the pathways connecting three QDs as depicted in Fig. \ref{fig:reducedmodel}(c). The PMF flux is calculated over the triangular area $S$:
\begin{align}
    \phi_{PMF}=\frac{e}{h}\oint_{S}B_{z}dA.
\end{align}
Since the QD locations are changed by the geometric deformation of the nanobubble, the enclosed area is accordingly changed by $\sigma_{x}$. Indeed, Fig. \ref{fig:reducedmodel}(d) shows the $\sigma_{x}$-dependent $\phi_{PMF}$, and the phase shift due to the $\sigma_{x}$ deformation is found to be about $\pi/4$. The transport characteristics corresponding to this $\pi/4$ phase shift are highlighted by a yellow box in Fig. \ref{fig:reducedmodel}(b), which is similar to the main finding of the manuscript.

\bibliographystyle{apsrev4-2}
\bibliography{EllipticNBQubit_abb}

\end{document}